\documentclass[apjl]{emulateapj} 

\usepackage{amsmath}  
\usepackage{amsfonts} 
\usepackage{graphicx} 
\usepackage{wasysym}
\usepackage{calc}
\usepackage{amssymb}
\usepackage{relsize}
\usepackage{natbib} 

\begin{document}


\title{Measuring Temporal Photon Bunching in Blackbody Radiation}


\author{P. K. Tan\altaffilmark{1,2,*}}
\author{G. H. Yeo\altaffilmark{2}}
\author{H. S. Poh\altaffilmark{1}}
\author{A. H. Chan\altaffilmark{2}}
\author{C. Kurtsiefer\altaffilmark{1,2,+}}
\altaffiltext{1}{Center for Quantum Technologies, 3 Science Drive 2, Singapore 117543}
\altaffiltext{2}{Department of Physics, National University of Singapore,
  Singapore 117551}
\altaffiltext{*}{email: pengkian@physics.org}
\altaffiltext{+}{email: phyck@nus.edu.sg} 



\begin{abstract}
Light from thermal black body radiators such as stars exhibits photon bunching
behaviour at sufficiently short time-scales. However, with available detector
bandwidths, this bunching signal is difficult to be directly used for intensity
interferometry with sufficient statistics in astronomy. Here we present an 
experimental technique to increase the photon bunching signal in blackbody
radiation via spectral filtering of the light source. Our measurements reveal
strong temporal photon bunching in light from blackbody radiation, including
the Sun. Such filtering techniques may revive the interest in intensity
interferometry as a tool in astronomy. 
\end{abstract}


\keywords{instrumentation: interferometers --- radiation mechanisms: thermal --- stars: fundamental parameters --- Sun: fundamental parameters --- techniques: interferometric}



\section{Background}

Statistical intensity fluctuations from stationary thermal light sources
formed the basis of the landmark experiments by \citet{hbt:54, hbt:57}, 
where correlation measurements between a pair of
photodetection signals were used to assess both temporal
and spatial coherence properties of light fields. The explanation for these
fluctuations as photon bunching phenomenon, as compared to a Poissonian
distribution for uncorrelated detection processes was a
major contribution for the development of the field of quantum optics \citep{glauber:63} 

\citet{hbt:56, hbt:74} developed their observation into a spatial intensity
interferometer which allowed them to infer stellar angular radii
from the coherence properties of star light over relatively large distances of 188\,m. Since a key quality of any telescope system is its angular resolution, given by its real or virtual aperture, they managed to
carry out these measurements with an unprecedented accuracy.

While stellar amplitude interferometry has evolved to become a standard technique to
increase the effective angular resolution of radio telescope arrays, the
angular resolution of the Hanbury-Brown--Twiss approach is still unmatched by
conventional diffraction-limited optical telescopes. It is also very
challenging to reproduce with Michelson interferometry due to the required
sub-wavelength optical path length stability \citep{davis:99}. Such setups involve remote locations for
observatory sites to reduce atmospheric turbulence, as well as expensive and
complicated adaptive optics to augment the measurable sensitivity by
suppressing optical noise \citep{millour:08}. Yet, the comparatively simple
optical intensity interferometry technique has not really found widespread
adoption in the observation of celestial objects. 

\pagebreak

\section{The problem}
Temporal photon bunching, i.e., the tendency for photons to cluster together
on a short time-scale, has been consistently observed from narrow-band
quasi-thermal light sources such as low pressure glow discharge lamps
\citep{martienssen:64, morgan:66}, and 
laser beams \citep{asakura:70} which are passed through a rotating ground
glass to randomize the optical phase in time.

Photon bunching in time for stationary fields is quantified by a second order
correlation function $g^{(2)}(\tau)$, which is the probability of observing a
second photodetection event at a time $\tau$ after a first one, normalized to the
probability of an uncorrelated detection. Optical modes in thermal states are
the textbook example where $g^{(2)}=2$ \citep{glauber:63, mandel:95,
  salehteich:07}. Blackbody radiation would 
therefore be the obvious choice for observing photon bunching, if it were
not for the short coherence time associated with its spectral properties.

In a scenario where spatial coherence effects are neglected, the second order
temporal correlation function for light fields can be written as
\begin{equation}
g^{(2)}(\tau)=1+\left|\gamma(\tau)\right|^2\,,
\end{equation}
where $\gamma(\tau)$ is the complex degree of coherence, or normalized
self-coherence of the stationary light field with $\gamma(0)=1$, which itself is proportional to
the Fourier transform of the spectral density $S(\nu)$ of the light field
\citep{mandel:95}. For well-defined spectral densities and coherence
functions, a meaningful coherence time $\tau_c$ can be defined, which is
reciprocal to a characteristic width $\Delta \nu$ of the spectral distribution
$S(\nu)$. 

For blackbody light sources in the optical regime, the coherence time
$\tau_{c}$ given by Planck's law is on the order of $10^{-14}$\,s, which is much
shorter than the best timing resolution $\tau_{t}$ of a few
$10^{-11}$\,s of existing photodetectors. 
As shown by \citet{scarl:68}, the observable excess from uncorrelated
pair detections, $g^{(2)}(\tau = 0)-1$, scales as $\tau_{c}/\tau_{t}$
and is thus only significant when the coherence time of the
source is of the order of the detection timing response.

Therefore, empirical observation of temporal photon bunching from blackbody
light sources has been extremely difficult, with only recent breakthroughs
using two-photon absorption in semiconductors \citep{boitier:09}, and in
ghost imaging research using a narrowband Faraday anomalous dispersion optical
filter \citep{karmakar:12, wu:14}.

In this paper, we present a spectral filtering technique that significantly
increases the the observed photon bunching signature from a blackbody light
source, such that intensity interferometry techniques for
observing celestial objects may become attractive again.

\section{Experiment}
\begin{figure}
\centering
\includegraphics[width=\hsize]{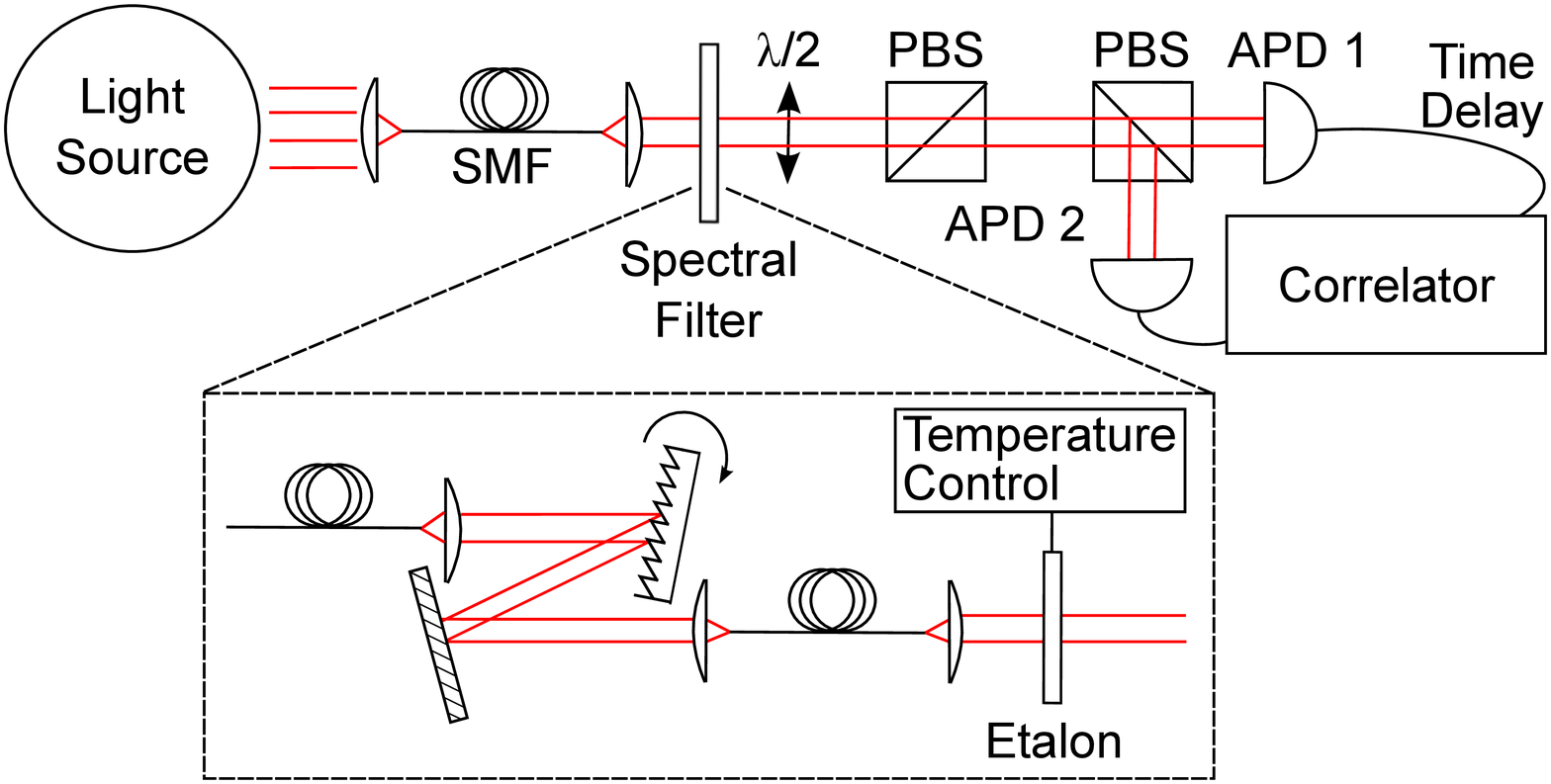}
\caption{Setup to determine the temporal correlation $g^{{(2)}}(\tau)$ for
  wideband thermal light. The initial spectrum gets filtered to a narrow
  optical bandwidth such that the temporal decay of the second order
  correlation function can be observed with conventional single photon
  detectors in a Hanbury-Brown--Twiss experiment. We employ a combination of
  a grating and a temperature-tuned etalon as a spectral filter, and ensure
  spatial coherence in the setup by using single mode optical fibers (SMF).}
\label{fig:setup}
\end{figure}

As with previous attempts to observe photon bunching from blackbody radiation
\citep{karmakar:12, wu:14}, we employ narrowband optical filtering of the
light source in order to increase 
the coherence time $\tau_c$, and use single photon detectors with low
timing uncertainties. The experimental setup is shown in
Fig.~\ref{fig:setup}.

Firstly, light from different sources under investigation is coupled into a
single mode optical fiber to enforce spatial coherence, because the
photon bunching signature $g^{(2)}(\tau=0)-1\propto1/M$, where $M$ is the
number of transverse modes \citep{glauber:63}. 

The light is then collimated with an aspheric lens and sent through the first
spectral filter, which is either an interferometric bandpass filter with a full width at half maximum (FWHM) of 3\,nm centered at 546.1\,nm to match the
$6p7s\,^3S_1 - 6p6s\,^3P^0_2$ transition in a Mercury (Hg) discharge lamp, or a grating
monochromator for broadband continuous spectrum. While the interference filter
is sufficient to select the spectral lines of the different mercury isotopes,
the grating monochromator had to be used for the blackbody sources.

\begin{figure}
\centering
\includegraphics[width=\hsize]{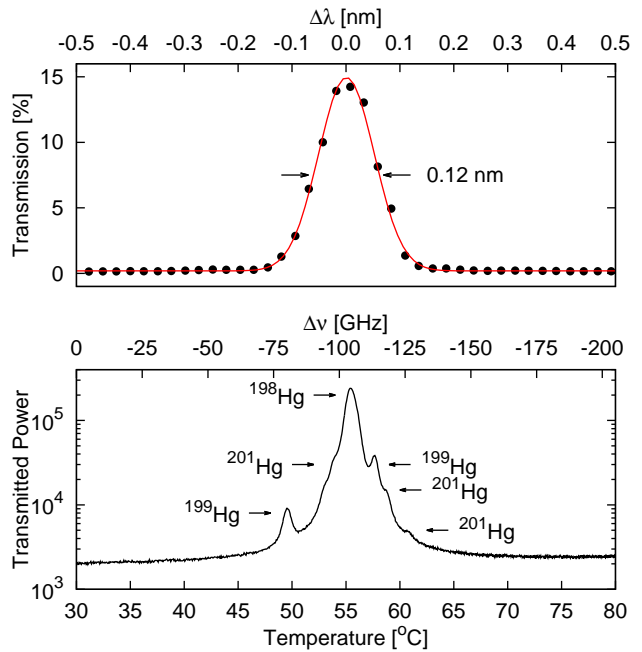}
\caption{(Top) Transmission profile of grating monochromator. The solid line
  is a fit to a Gaussian profile to measurements for different grating angles
  (symbols), resulting in a full width at half maximum (FWHM) of
  0.122$\pm$0.002\,nm. (Bottom) Transmission of Hg light near 546\,nm through the
  temperature-tuned etalon with 2\,GHz bandwidth. The hyperfine structure for
  different isotopes is partially resolved, as identified in
  \citep{sansonetti:10}.}
\label{fig:transmission}
\end{figure}

The core of the monochromator is a blazed reflective diffraction grating
with 1200\,lines/mm mounted on a rotation stage. Light from the
optical fiber (NA = 0.13) is collimated with an achromatic doublet ($f=30\,$mm),
illuminating about $10^4$ lines of the grating. The first diffraction order is
imaged with another achromat of ($f=30\,$mm) into another single mode optical
fiber. This arrangement acts as a tunable narrowband filter in the
visible regime, allowing to adjust for peak emissions
of blackbody light sources. A test of the transmission of a narrowband laser
source ($\lambda$ = 532\,nm) is shown in Fig.~\ref{fig:transmission} (top
part). The fit to a Gaussian transmission profile leads to an optical
bandwidth of 0.122$\pm$0.002\,nm (FWHM). We typically observe a transmission
of narrowband light from single mode fiber to single mode fiber of
$\approx$15\%.

In a second step, the light passed through a solid etalon (thickness
$d$=0.5\,mm) with flat surfaces. The etalon has a free spectral range of
$FSR=c/(2nd)=$ 205\,GHz, with a refractive index $n=1.46$ for the fused
silica material (Suprasil311), and the speed of light $c$ in vacuum. At
$\lambda$ = 546\,nm, the free spectral range corresponds to
$\Delta\lambda=\lambda^2/c\cdot FSR\approx$ 0.2\,nm, so that only one
transmission order should fall within the transmission width of the
monochromator.
Both surfaces of the etalon have dielectric coatings with a
reflectivity of $R\approx$ 97\% over a wavelength range from 390\,nm to
810\,nm, resulting in a transmission bandwidth (FWHM) of
\begin{equation}
  \Delta\nu \approx FSR\cdot{1-R\over\pi\sqrt{R}}\approx2\,\mathrm{GHz}\,.
\end{equation}

The central transmission frequency of the etalon is tuned by temperature, with
a scaling rate of -4.1\,GHz/K due to thermal expansion and temperature
dependence of the refractive index. A temperature stability of $\pm$5\,mK over the
measurement duration of several hours ensures that the etalon stays on
a specific wavelength. Temperature tuning allows the etalon to be always used
under normal incidence, which avoids frequency walk-off
\citep{green:80} and spatial distortion \citep{park:05}. To demonstrate the
resolving power of this device, we tuned the peak transmission across the
various spectral lines from a Hg gas discharge lamp at 546.1\,nm, revealing a
partially resolved hyperfine structure for different isotopes
(see Fig.~\ref{fig:transmission}, bottom part).

A Glan-Taylor polarizer (PBS) following the etalon selects
a well-defined linear polarization mode for the photons, with the same
rationale of minimizing the number of modes ending up in the temporal
correlation setup. A half wave plate ($\lambda/2$) transforms the
polarization emerging from the grating/fiber combination such that there is
maximal transmission through the PBS. 

A second Glan-Taylor polarizer, rotated by $\approx45^\circ$  with respect to
the polarization direction of the first, distributes the light onto two
silicon avalanche photodetectors (APD) that provide detection timing signals
to a correlator. The relative orientation of the two polarizers allows for the
balancing of the photodetection rates to optimize the integration time, and
also helps to suppress  the optical crosstalk 
between the two APDs due to hot carrier recombination in the devices upon
detection which would otherwise introduce false coincidence events
\citep{kurtsiefer:01a}.

The temporal correlation was carried out with a digital sampling
oscilloscope (4\,GHz analog bandwidth, 40\,Gsamples/sec, LeCroy) that can
measure and histogram detection timing differences between two APD events with a
timing uncertainty below 10\,ps. 

\begin{figure}
\centerline{\includegraphics[width=\hsize]{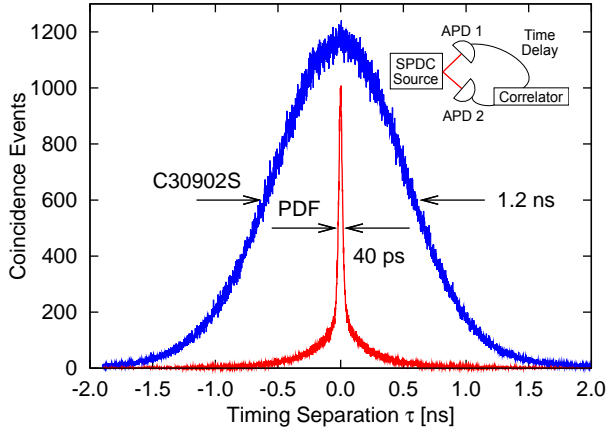}}
\caption{Comparison of timing jitter between two actively quenched thin avalanche
  photodiodes (PDF), and two passively quenched deep avalanche photodiodes
  (C90302S) by photopair detection from light generated by spontaneous
  parametric down conversion (SPDC) with a coherence time below 10\,ps.}
\label{fig:jitter}
\end{figure}

To assess the overall timing uncertainty with the detector/correlator combination,
we exposed two different avalanche photodetector sets to photon pairs
generated by spontaneuous parametric down conversion (SPDC) with a coherence time below
10\,ps. The resulting temporal correlations shown in
Fig.~\ref{fig:jitter} reflects the overall timing resolution of the system.
For APDs with a deep conversion zone (C30902S, Perkin-Elmer, passive
quenching), the coincidence distribution appears roughly Gaussian, with 1.2\,ns
FWHM. Since the photon bunching signature $g^{(2)}(\tau)$ of the light field
under investigation needs to be convoluted with this detector response, we
would observe a significant reduction of the photon bunching peak with
these detectors. A second set of APDs, based on thin conversion zones (type
PDF, Micro Photon Devices, active quenching), leads to a much better localized
coincidence distribution with 40\,ps FWHM, although a significant fraction of
the distribution shows a tail with timing differences up to 500\,ps.

\section{Results}

\begin{figure}
\centering
\includegraphics[width=\hsize]{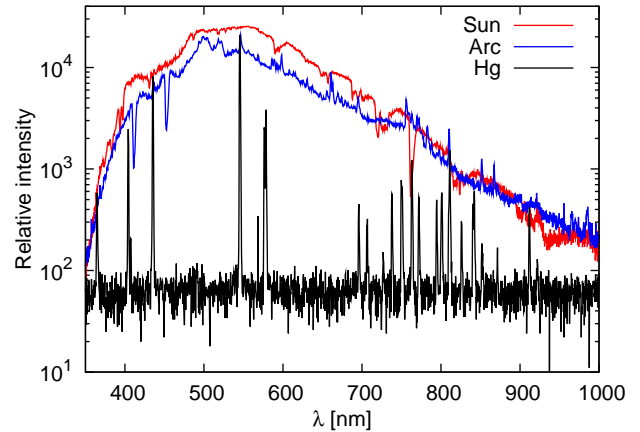}
\caption{Spectrum of a Hg glow discharge lamp, an Ar arc lamp as a simulator of a stellar light source, and the Sun.}
\label{fig:spectrum}
\end{figure}

\begin{figure}
\centering
\includegraphics[width=\hsize]{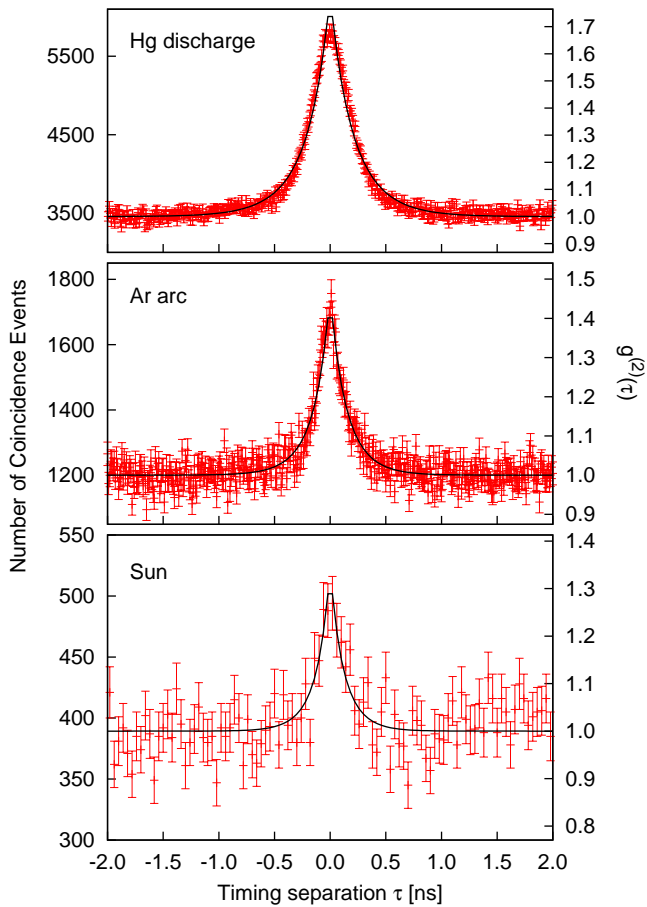}
\caption{Intensity correlation function $g^{{(2)}}(\tau)$ from the three
  thermal light sources after spectral filtering. All light sources show
  a significant bunching at $\tau=0$, and a decay time that is
  well-resolved with the avalanche photodetectors. Error bars indicate
  uncertainties due to Poissonian counting statistics.}
\label{fig:g2}
\end{figure}

Measurements for $g^{{(2)}}(\tau)$ were performed on three light sources: a
low pressure Mercury (Hg) discharge lamp as a quasi-thermal light source, a
radio frequency-driven Argon (Ar) arc discharge lamp to produce a plasma with a
blackbody temperature $\approx$ 6000\,K as the thermal light source, and the
Sun as a celestial thermal light source. Their respective spectral distributions are shown in Fig.~\ref{fig:spectrum}. The optical filters were fixed to
maximize transmission for the spectral lines around 546.1\,nm in the Hg
spectrum. The resulting timing difference histograms are shown in
Fig.~\ref{fig:g2}.

To ensure proper normalization of the resulting 
$g^{(2)}(\tau)$, we independently recorded pair events  with
a resolution of about 300\,ps with a time stamp unit that, unlike the
oscilloscope, does not exhibit a significant dead time, and found that
$g^{(2)}(|\tau|\gtrsim2\,\textrm{ns})\rightarrow1$. To extract a meaningful
set of model parameters, we assumed a Lorentzian line shape for the spectral
power distribution; this is mostly to reflect the approximate transmission
function of the etalon. A corresponding model function for the obtained pair
rates in the histograms is
\begin{equation}
N(\tau)=a+b\cdot e^{-\left|2\tau/\tau_c\right|}\,
\end{equation}
where $a,b$ capture the absolute event number and the degree of bunching,
respectively, while $\tau_c$ represents a coherence time. The $g^{(2)}$ scale
on the right side of diagrams in Fig.~\ref{fig:g2} refers to the absolute
number of events after division by fit parameter $a$. 

For the Hg emission line at 546\,nm from the low pressure glow discharge lamp,
coincidence data was obtained over an
integration time of 68 hours, with single detector rates of 115000\,s$^{-1}$.
The relatively large discrepancy between expected coincidences with the single
rates and the observed coincidences is due to a dead time of the oscilloscope
in this measurement.   
From the numerical fit, we find $g^{{(2)}}(\tau=0)=1.79\pm0.01$
and an exponential decay with a characteristic coherence time of $\tau_{{c}}$
= 0.436$\pm$0.006\,ns. The photon bunching signature comes very close to the
expected value of 2 for a thermal state, with the difference possibly due to
the residual timing uncertainty of our photodetectors.
While the Doppler-broadened line width for individual transitions in
Hg vapour at temperatures around 400\,K is on the order of 235\,MHz, the
contributions of different hyperfine transitions from different isotopes
result in a complex overall line shape of several GHz width, out of
which we select a reasonable fraction with the etalon. 

For the Ar arc discharge lamp, a wavelength of 540\,nm was
chosen to avoid contributions of a prominent peak at 546\,nm, possibly caused by mercury traces in the plasma. We find that
$g^{{(2)}}(0) =1.45\pm0.03$ and a shorter coherence time
$\tau_{c}=0.31\pm0.01$\,ns with a reduced $\chi^{2}_{r}=1.066$.
The wider spectrum of the cut-out blackbody radiation makes the decay time
shorter, so a peak reduction due to the detector response becomes more
prominent. Furthermore, residual transmission of the grating monochromator or
stray light into other transmission peaks of the etalon may contribute to
light that is not coherent with the main peak on observable time-scales. 

We finally collected sunlight into a single mode optical fiber with a
coelostat and transferred it into the lab. At the same working wavelength of
546.1\,nm, we find $g^{{(2)}}(0)=1.3\pm0.1$ and a coherence
time $\tau_{c}= 0.26\pm0.05$\,ns, with $\chi^{2}_{r}=1.0244$ over an
integration time of 45 minutes. At the moment, we cannot explain
the reduction of the visibility of the photon
bunching, but seeing-induced atmospheric fluctuations in optical path
length differences \citep{dravins:07a} may be a reason.

\section{Summary and Outlook}

We have successfully developed and demonstrated an experimental technique to
resolve and explicitly measure the temporal photon bunching from thermal
blackbody light sources, including both discrete spectrum and blackbody
spectrum such as the Sun. We observed a photon bunching signature, i.e., the
excess of $g^{(2)}(0)$ above 1, that exceeds the recently reported records
of $g^{(2)}(0)=1.03$ \citep{karmakar:12} and $g^{(2)}(0)=1.04$ \citep{wu:14}
by about an order of magnitude. 

Compared to photon bunching observations using two-photon absorption
\citep{boitier:09}, our method would work at astronomical
intensity levels, and not be constrained to cases with the simultaneous arrivals of two photons at
the same detector.

We believe that with better timing control of the photodetectors and a better
match of the etalon with the grating monochromator as a first stage, it should
not be too technically challenging to reach the theoretical limit $g^{(2)}(0)=2$
for blackbody radiators.

The substantial increase in the observable photon bunching signature may
reopen considerations of intensity interferometry as a useful tool in
astronomical applications, where large baselines can be implemented with
significantly less effort than that in first order (amplitude) stellar
interferometers. Current atomic clock technology should 
even allow for the development of very large scale optical intensity
interferometers with kilometric baselines, leading to micro-arcsecond
resolution. With this, one could investigate stellar limb darkening coefficients, resolve spectroscopic binaries, measure stellar oblateness to better understand the internal dynamics, and also determine the pulsations of cepheid variables to better derive the Hubble constant \citep{barbieri:09, borra:13, dravins:07, dravins:13, foellmi:09, lebohec:08, naletto:09, nunez:12, ofir:06, rou:12, wittkowski:04}.

Beyond the traditional application of estimating the angular size of a
celestial object via its spatial coherence length, this technique may also be
used to measure photon decoherence of starlight across extended time-scales
and distances, which are impossible to create in Earth-based
laboratory conditions. Such experiments may provide complementary
observational evidence that help to set constraints in theoretical models of
quantum gravity \citep{lieu:03, maziashvili:09, ng:03, ragazzoni:03}. 



\acknowledgments
We acknowledge the support of this work by the National Research Foundation and the Ministry of Education in Singapore.

\end{document}